**Seismic regularity coefficient changes precede stronger earthquakes in Santorini swarm**

S. Lasocki[1], B. Orlecka-Sikora[1], A. Kostoglou[1], V. Karakostas[2], E. Papadimitriou[2]

[1]Institute of Geophysics, Polish Academy of Sciences

[2]Geophysics Department, Aristotle University of Thessaloniki

**Abstract:**

We have been analysing the ongoing seismic swarm in Santorini, Greece, with the Seismic Regularity ($d_c$) and Seismic Strain Dynamics (*SSD*) coefficients, which quantify temporal changes in seismic process organization and strain energy release. Stronger earthquakes are preceded by characteristic down-up changes of $d_c$. Such a change indicates an increase of the seismic process regularity followed by a fast transition to a more irregular and chaotic seismic regime.

**Main Text**

Since 27 January 2025, an intense seismic swarm has occurred close to the Greek island of Santorini. Until 13 February 10:45, the National Observatory of Athens (NOA) seismic network recorded close to 2,200 events, including twenty events of magnitude from 4.8 up. This activity has significantly impacted residents and tourists, prompting evacuations, school closures, and emergency response measures.

The Santorini-Amorgos region is seismically and volcanically active within the Hellenic subduction zone, where the African plate subducts beneath the Aegean plate at 0.9 cm/year[1]. It is a tectonically complex region, where extensional forces dominate, creating a network of faults and fractures that facilitate magma migration. The largest recorded events - the Amorgos earthquakes (Ms 7.4 and 7.3) that occurred in 1956, triggered a tsunami and caused widespread destruction[2].

Santorini hosts an active caldera, responsible for the Minoan eruption (~1613 BCE), one of the most significant eruptions in history[3]. The current swarm activity concentrates close to the Kolumbo volcano, located 7 km northeast of Santorini. The Kolumbo volcano last erupted in 1650 CE, triggering a submarine explosion and a tsunami[4].

We have investigated the dynamics of the ongoing swarm since 9 February with the Seismic Regularity ($d_c$) and Seismic Strain Dynamics (*SSD*) coefficients.

Let be a seismic series of *n* successive events, {$t_i$, $lat_i$, $lon_i$, $m_i$}, i=1,..,n, where *t* is the event occurrence time, *lat* and *lon* are the geographical coordinates of its epicenter, and *m* is its magnitude. To evaluate the Seismic Regularity coefficient, we build a new series {$t_i$, $dt_i$, $dr_i$, $m_i$}, i=2,..,n, where $dt_i=t_i-t_{i-1}$ is called the waiting time, and $dr_i$ is the orthodromic distance between epicenters of *i* and *i*-1 events called the epicentral offset. The *dt*, *dr*, and *m* parameters are transformed into equivalent dimensions[5]. The transforms *DT*, *DR*, and *MC* are comparable and have Euclidean metric in the [0,1] cube. For a prescribed data window containing *K* consecutive {$t_k$, $DT_k$, $DR_k$, $MC_k$}, k=1,..,K≤(n-1), the Seismic Regularity coefficient reads:

$$d_c(t_K) = \frac{2}{K(K-1)} \sum_{k=1}^{K-1} \sum_{j=k+1}^{K} \sqrt{(DT_k - DT_j)^2 + (DR_k - DR_j)^2 + (MC_k - MC_j)^2} \quad (1)$$

The seismic regularity coefficient tracks time changes in seismic process regularity. A decrease of $d_c$ indicates that the waiting times, epicentral offsets, and magnitudes of *K* events became similar; thence the seismic process was more regular. Conversely, an increase in $d_c$ indicates a transition toward irregularity, which could result from stress redistribution or increased complexity in rupture processes. Previous studies have shown that large earthquakes are often preceded by significant drops in $d_c$[6,7].

The Seismic Strain Dynamics (*SSD*) coefficient is computed from the $K$ consecutive events data window drawn from the initial seismic series:

$$SSD(t_K) = log\dot\varepsilon(t_K) = log\frac{\partial}{\partial t}\left(\sum_{i=1}^{K} M_{0,i}^{\zeta}(t_i)\right) = log\frac{\sum_{i=1}^{K} M_{0,i}^{\zeta}}{dt_i}, \qquad (2)$$

where $\dot\varepsilon(t)$ is the strain rate, $M_{0i}$ is the seismic moment of the $i$-th event, and $dt_i$ is the waiting time. The choice of $\zeta$ determines the relationship to specific strain proxies: $\zeta = 1/2$ corresponds to cumulative Benioff strain; $\zeta = 0$ refers to the seismic event rate[8]. For $\zeta = 1/3$, $SSD$ measures the velocity of Subcritical Fracture Growth (SFG), a process occurring under subcritical stress conditions[9]. We use local magnitude ($M_L$), from earthquake catalogues instead of the seismic moment to compute $SSD$ because both moment and magnitude have the same trends. Such an approach allows for practical application in cases where seismic moment estimates are unavailable while maintaining consistency with strain evolution analysis.

The Seismic Strain Dynamics coefficient quantifies the rate of cumulative seismic strain, providing insights into the progressive damage accumulation within a fault system. A decrease in *SSD* indicates a stabilization phase with lower strain release rates, whereas its increase signals enhanced strain accumulation, potentially foreshadowing failure. Previous studies have linked *SSD* trends to seismic preparatory phases[7], revealing characteristic strain acceleration before major earthquakes.

We were using the P and S phases from NOA. Thirty-four stations with distances of up to 250 km were chosen, achieving good azimuthal coverage. We applied the 1D crustal velocity model for the Anydros basin[10] and the Vp/Vs ratio equal to 1.73, using earthquakes with more than ten S phase readings. We calculated station delays to account for lateral crustal variations by applying the iterative procedure[11] with the use of the HYPOINVERSE program[12]. The initial relocation was done with HYPOINVERSE, using the local model and station delays. The final relocation was performed with the hypoDD[13], applying 25 iterations (5 sets of 5 iterations). The differential times were calculated from the event phases times. The magnitude completeness level, calculated using the modified Goodness-of-Fit Test[14], was 3.2.

The Santorini swarm activity has been continuing, providing new seismic data. We have been repeating the analysis about once a day, each time with the grown input catalogue. The results presented here were obtained for the catalogue until 12 February 19:59.

Figure 1 presents the time changes of the seismic regularity coefficient, $d_c$ calculated from the complete part of the catalogue using the 50-element data window being shifted of one element. Stars and circles by the top of the graph indicate occurrence times and magnitudes of $M \geq 4.8$ earthquakes. The correlation between the $d_c$ fluctuations and strong event occurrences is very distinct. Most of the strong events are preceded by a drop in $d_c$ followed by its more or less developed increase. This picture suggests that there was a transition toward a more regular seismic process before stronger earthquakes, potentially linked to stress accumulation and fault synchronization. The afterward rises of $d_c$ illustrate returns to a more chaotic seismic regime, likely linked to a stress build-up and an enhanced development of process zones of smaller faults before the next strong event.

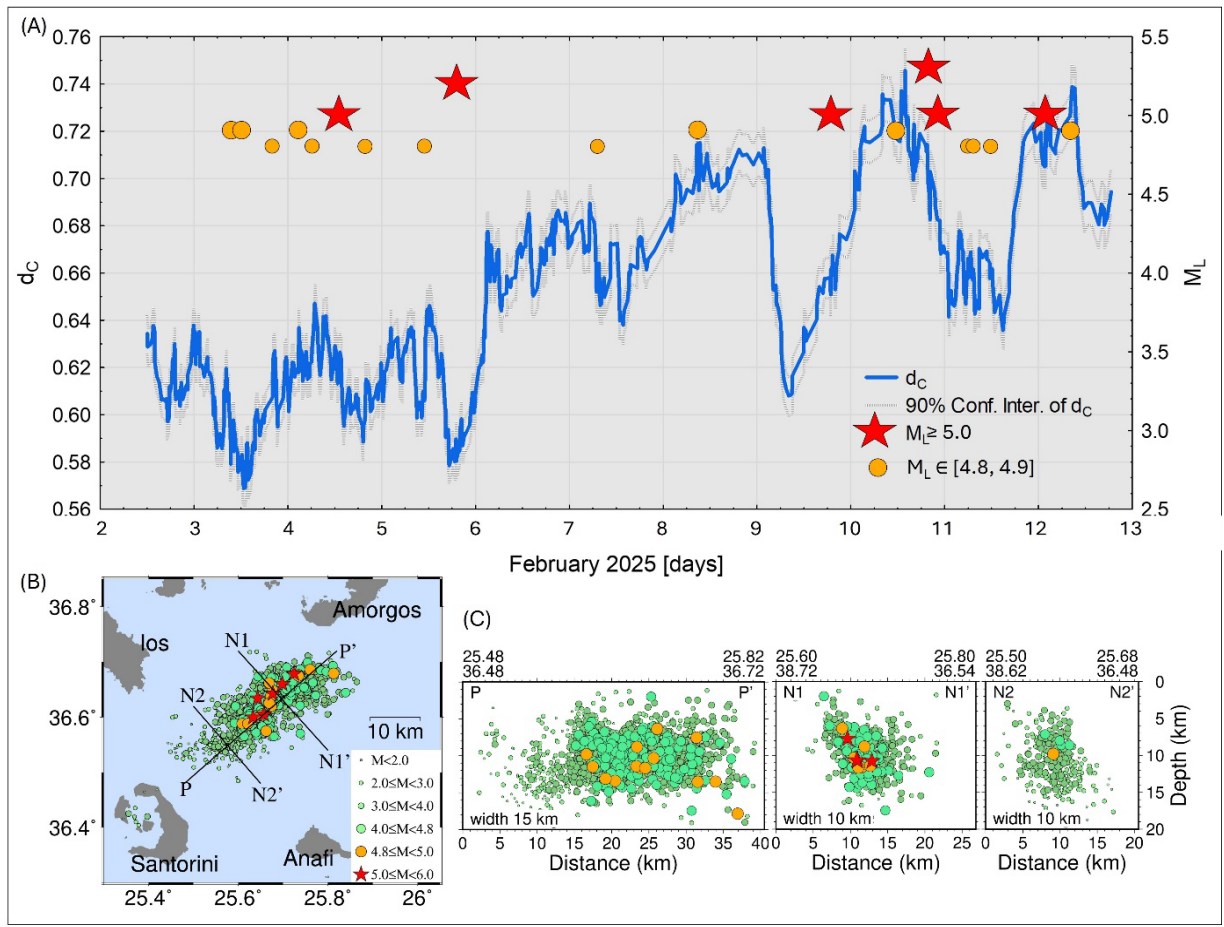

*Figure 1. A – time-variations of the Seismic Regularity coefficient, $d_c$. Stars and circles by the graph top indicate the occurrence of stronger earthquakes. B and C – surface and depth sections of the swarm location.*

The portrait of Santorini seismic process changes obtained from the $d_c$ time-variations (Figure 1) is complemented in Figure 2, where the time changes of the Seismic Strain Dynamics coefficient are exposed. The behaviour of *SSD* for the Santorini swarm is strikingly similar to the variations of *SSD* for acoustic emission during the progressive damage accumulation and strain localization leading up to the failure of Westerly Granite samples in the laboratory stick-slip experiments[15]. It seems that the first damage cycle may have ended with the so-far strongest earthquake (M5.3). The *SSD* picture afterward, specifically the stabilization of *SSD* at some 2.4 suggests that the next stress accumulation began, which may lead to another strong event.

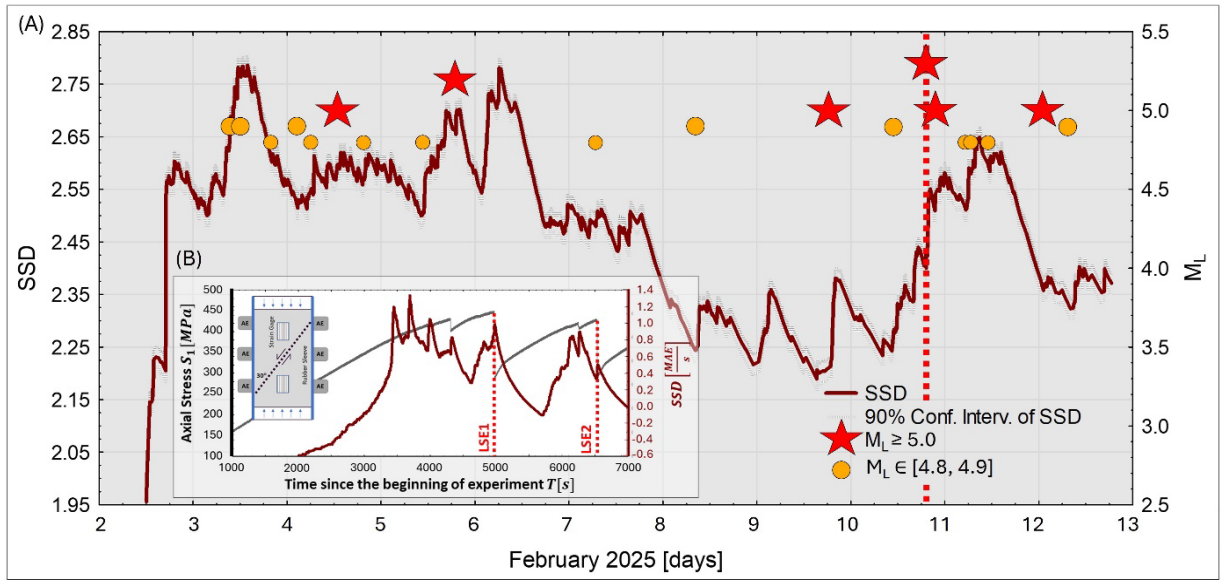

*Figure 2. A – time-variations of the Seismic Strain Dynamics coefficient, SSD. Stars and circles by the graph top indicate the occurrence of stronger earthquakes. The red vertical line located at the strongest earthquake (M5.3) occurrence time marks the suggested end of the first damage cycle. B – time-variations of SSD during Westerly Granite laboratory experiment[14]. LSEx mark the large slip events during the loading cycles.*

Since 11 February, after we completed the analysis shown in Figures 1 and 2 the Santorini swarm activity distinctly decreased (Figure 3). Based on that, it was expected that this activity would die down soon. However, these expectations were not confirmed by the behaviour of the Seismic Regularity coefficient, $d_c$, which did not stabilize but was constantly going down. After it reached the lowest level on 16 February, equal to 0.57, it moved up, as shown in Figure 4. This newest down-up behaviour of $d_c$ preceded the next strong earthquake (M5.1) that occurred on 17 February at 7:49:50.92.

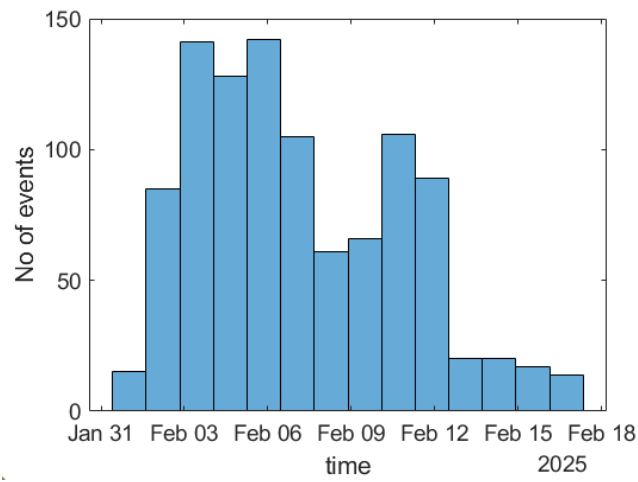

*Figure 3. Seismic activity of Santorini swarm. The histogram includes only events from the complete part of the catalogue (M≥3.2)*

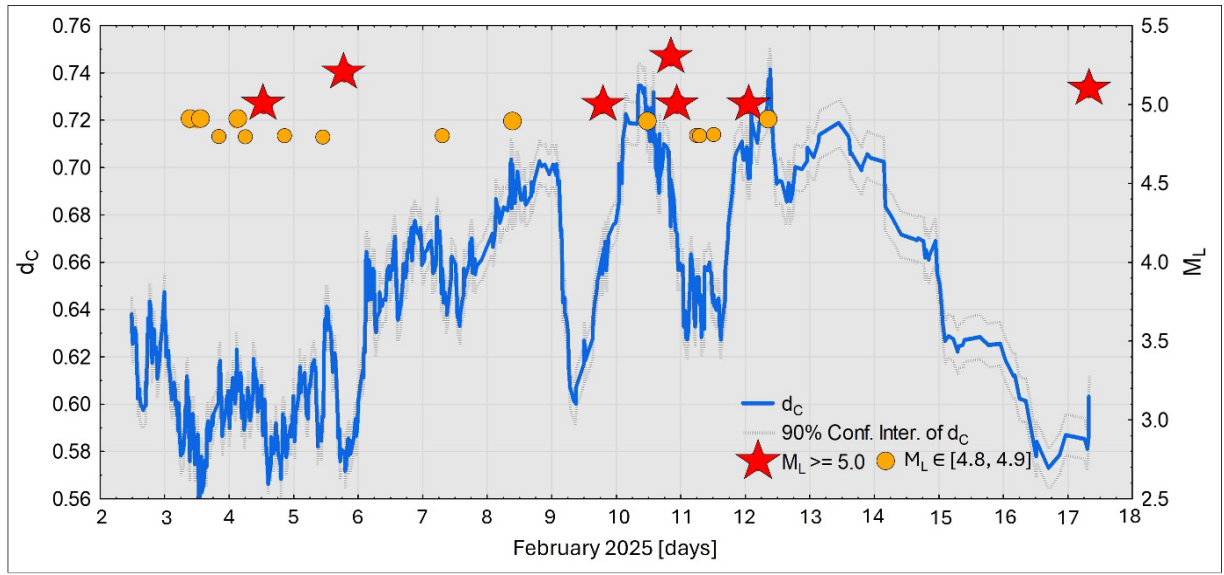

*Figure 4. Time-variations of the Seismic Regularity coefficient, dc. These results were obtained from the data gathered after the end of the analysis whose results are shown in Figures 1 and 2. Stars and circles by the graph top indicate the occurrence of stronger earthquakes.*

**References**

1. Reilinger R. et al. GPS constraints on continental deformation in the Africa-Arabia-Eurasia continental collision zone and implications for plate dynamics. *J Geophys Res.* **111**(B5), B05411 (2006).
2. Makropoulos K. C., Drakopoulos J. K., Latousakis J. B. A revised and extended earthquake catalogue for Greece since 1900  *Geophys J Int.* **98** (2), 391–394 (1989)
3. Friedrich, W. L. The Minoan eruption of Santorini around 1613 BCE and its consequences. *Antiquity* **87**(336), 54-65 (2013)
4. Dominey-Howes, D., G. Papadopoulos, Dawson A. Geological and historical investigation of the 1650 Mt. Columbo (Thera Island) eruption and tsunami, Aegean Sea, Greece. *Nat Hazards* **21**(1), 83–96 (2000).
5. Lasocki S. Transformation to equivalent dimensions – a new methodology to study earthquake clustering. *Geophys J Int.*, **197** (2), 1224-1235 (2014).
6. Lasocki S. et al. Premonitory earthquakes clustering process in an equivalent dimensions space before the 2017Mw 8.2 Tehuantepec, Mexico, mainshock. *Seismol Res Lett*. **96**, 340–352 (2024).
7. Orlecka-Sikora B., Ciechowska H, do Nascimento A.F. Insights into Dynamics of Earthquake Preparation Process in Water Reservoir-Triggered Seismic Zones, *J. Geol. Soc. India* (2025), accepted
8. Main, I.G. Applicability of time-to-failure analysis to accelerated strain before earthquakes and volcanic eruptions. *Geophys J Int.* **139**, F1–F6 (1999).
9. Atkinson, B.K. Subcritical crack growth in geological materials. *J Geophys Res.* **89**, 4077–4114 (1984).

10. Papadimitriou, P. et al. The Santorini Volcanic Complex: A detailed multi-parameter seismological approach with emphasis on the 2011–2012 unrest period. *J Geodyn.* **85**, 32–57 (2015).

11. Karakostas, V., Karagianni, E., Paradisopoulou, P. Space-time analysis, faulting and triggering of the 2010 earthquake doublet in western Corinth Gulf. *Nat Hazards* **63**, 1181–1202 (2012).

12. Klein, Fred.W. User's Guide to HYPOINVERSE-2000, a Fortran Program to Solve for Earthquake Locations and Magnitudes. *U.S. Geol. Surv. Open File Report* 02-171 (2002).doi:http://geopubs.wr.usgs.gov/open-file/of02-171/

13. Waldhauser, F. hypoDD - A Program to Compute Double-Difference Hypocenter Locations, *US Geol. Surv. Open File Rep.* 1–25 (2001). doi:http://geopubs.wr.usgs.gov/open-file/of01-113/

14. Leptokaropoulos, K.M., V.G. Karakostas, E.E. Papadimitriou, A.K. Adamaki, O. Tan, and S. Inan (2013). A homogeneous earthquake catalog for Western Turkey and magnitude of completeness determination, Bull. Seism. Soc. Am. 103(5), 2739–2751

15. Kwiatek, G.; Goebel, T. (2024): Acoustic Emission and Seismic moment tensor catalogs associated with triaxial stick-slip experiments performed on Westerly Granite samples. GFZ Data Services. https://doi.org/10.5880/GFZ.4.2.2023.003

**Data availability**

The seismic phases used for this study were downloaded through the database of the National Observatory of Athens (NOA). https://bbnet.gein.noa.gr/Events/. Accessed 13 July 2025.